\documentclass[twocolumn,a4paper,amsmath,amssymb,showpacs,prb,superscriptaddress]{revtex4-2}

\usepackage[dvipdfmx]{graphicx}
\usepackage{bmpsize}
\usepackage{amssymb, amsfonts, amsmath}
\usepackage{dcolumn}
\usepackage{bm}
\usepackage{color}
\usepackage[normalem]{ulem}
\usepackage{braket} 

\newcommand{\red}{\color{black}}

\begin{document}
\title{High-performance descriptor for magnetic materials:\newline
Accurate discrimination of magnetic \red{structure}}
\author{Michi-To Suzuki}
\affiliation{Center for Computational Materials Science, Institute for
Materials Research, Tohoku University, Sendai, Miyagi 980-8577, Japan}
\affiliation{Center for Spintronics Research Network, Graduate School of
Engineering Science, Osaka University, Toyonaka, Osaka 560-8531, Japan}
\author{Takuya Nomoto}
\affiliation{Research Center for Advanced Science and Technology, University of Tokyo, Komaba, Meguro-ku, Tokyo 153-8904, Japan}
\author{Eiaki V. Morooka}
\affiliation{Department of Applied Physics, Aalto University, P.O. Box 11100, 00076 Aalto, Espoo, Finland}
\author{Yuki Yanagi}
\affiliation{Liberal Arts and Sciences, Toyama Prefectural University, Imizu, Toyama 939-0398, Japan}
\author{Hiroaki Kusunose}
\affiliation{Department of Physics, Meiji University, Kawasaki 214-8571, Japan}
\date{\today}

\begin{abstract}
The magnetic structure is crucial in determining the physical properties
 inherent in magnetic compounds. We present an adequate descriptor for
 magnetic structure with proper magnetic symmetry and high
 discrimination performance, which does not depend on artificial choices
 for coordinate origin, axis, and magnetic unit cell in crystal.
  We extend the formalism called ``smooth overlap of atomic positions''
 (SOAP), providing a numerical representation of atomic configurations to 
that of magnetic moment configurations. We introduce the descriptor in
 terms of the vector spherical harmonics to describe a magnetic moment
 configuration and partial spectra from the expansion coefficients.
We discuss that the lowest-order partial spectrum is insufficient to discriminate 
the magnetic structures with different magnetic anisotropy, and a
 higher-order partial spectrum is required in general to differentiate
 detailed magnetic structures on the same atomic configuration.
We then introduce the fourth-order partial spectrum and evaluate the
 discrimination performance for different magnetic structures, mainly
 focusing on the difference in magnetic symmetry.
The modified partial spectra that are defined not to reflect the difference of magnetic anisotropy
are also useful in evaluating magnetic structures obtained from the first-principles calculations
 performed without spin-orbit coupling.
We apply the present method to the symmetry-classified magnetic
 structures for the crystals of Mn$_3$Ir and Mn$_3$Sn, which are known to exhibit anomalous transport
 under the antiferromagnetic order, and examine the discrimination performance of 
the descriptor for different magnetic structures on the same crystal.
\end{abstract}

\maketitle

\section{INTRODUCTION}
The recent significant advance in research using machine-learning
techniques in quantum chemistry and condensed matter physics gets
benefits from the development of high-performance descriptors that transform
the structural information of molecules and crystals to data-style
representations friendly to machine-learning applications.
Various descriptors have been proposed for atomic configurations of
molecule{\red s} and crystal{\red s}~\cite{Behler2007,Bartok2010,Jorg2011,Rupp2012,Bartok2013,Faber2015,Glielmo2017,Huo2018}
and applied to the analysis of nonmagnetic materials~\cite{Bartok2018, Fujii2020}.

For magnetic compounds, magnetic structure is crucial in determining their
physical properties, and it must be taken into account as additional
degrees of freedom in the descriptor.
So far, only a few descriptors have been proposed to describe magnetic
compounds, including the information on magnetic moments on each atomic site.
``Moment Tensor Potentials{\red ,}'' ``atomic symmetry functions{\red ,}'' and ``Smooth Overlap
of Atomic Positions (SOAP)'' {\red have been} recently developed to encode the characters of magnetic
materials~\cite{Eckhoff2021,Novikov2022,Domina2022}. 
These descriptors are applicable to distinguish different collinear
magnetic structures, whose spin moments on each atomic site are parallel
or anti-parallel, having either up or down-spin moments.
However, in the more general cases including non-collinear magnetic structures,
the descriptors must properly preserve the information on the directions
of magnetic moments to analyze the macroscopic magnetic properties.
 The anomalous/topological Hall effect, nonreciprocal charge transport,
 and magnetoelectric effect are such examples, which are characterized
 by magnetic symmetry according to Neumann's principle.


Domina et al. discuss a straightforward extension of the power spectrum
 of SOAP~\cite{Bartok2013}, in which the multi-dimensional vector is
 used to express atomic positions from the expansion coefficients of
 spherical harmonics for atomic density{\red ,} and the power spectrum is calculated for
magnetic configurations on atomic clusters~\cite{Domina2022}.
However, as discussed in the present paper, the power spectrum calculated from the
 magnetization density is {\red insufficient} to characterize the magnetic
 structures on the high{\red -}symmetry atomic systems{\red ,} and {a}
 higher{\red -}order
 partial spectrum is required to distinguish the magnetic structures
 with different magnetic symmetries.
In this paper, we discuss four types of partial spectra, i.e.,
 second and fourth-order partial spectra referred {\red to} as {\red power spectrum and trispectrum}, respectively, and those modified to neglect the
 difference of magnetic anisotropy
 {\red in} the magnetic structures and investigate the discrimination performance of
 those partial spectra for the high-symmetry magnetic structures.
We also show that the modified partial spectra {\red are} useful {\red
 in classifying} the magnetic structures obtained from the
 first-principles calculations without considering the spin-orbit
 coupling.

The paper is organized as follows. 
Section~\ref{Sec:Formulation} provides the formulation to transform the
magnetic structure to partial spectra.
In Sec.~\ref{Sec:MultipoleExp}, we define the magnetization density
representing magnetic structures and provide the multipole expansion
of the magnetization density with the explicit form of the expansion
coefficients, as discussed in the literature~\cite{Domina2022}.
In Sec.~\ref{Sec:SpectraSO}, {\red w}e derive the partial spectra from
the {\red second and higher-}order similarity kernel{\red s} defined by
the overlap integral of different magnetic environments.
In Sec.~\ref{Sec:SpectraNSO}, we discuss that the modified partial
spectra that are defined not {\red reflecting} the difference {\red in} magnetic anisotropy. 
 Section~\ref{Sec:NumericalTest} discusses the discrimination performance of
 the magnetic structures on the same atomic configurations. 
 In Sec.~\ref{Sec:ParameterTest}, we discuss the basic behavior of
 partial spectra for the parameters introduced in the calculations by
 using magnetic configurations on a simple one{\red -}dimensional crystal structure.
 In Sec.~\ref{Sec:SymmetryTest}, the method is applied for
 the symmetrized magnetic structures classified according to the
 magnetic symmetries defined on the crystals of Mn$_3$Ir, which has
 a simple cubic structure, and Mn$_3$Sn, a hexagonal structure.
The numerical tests for the magnetic compounds provide the knowledge of
the appropriate choice of the derived magnetic partial spectra depending on
magnetic environments.
The new representation scheme of the magnetic moment configurations thus provides a
solid foundation for machine learning, which is applicable to the study of
magnetic materials, as summarized in Sec.~\ref{Sec:Summary}.


\section{Formulation}
\label{Sec:Formulation}

This section provides formulation{\red s for transforming} a given magnetic
 structure on the atomic cluster or crystal into a multi-dimensional vector.
 The outline of the procedure is, first, convert the magnetic moment configuration to
a magnetization density and, second, expand {\red the magnetization
 density by vector spherical harmonics with appropriate radial functions} and, finally,
construct the partial spectrum from the expansion coefficients in a
similar way to obtain the partial spectrum from the expansion
coefficients for the spherical harmonics expansion of the atomic density in
SOAP~\cite{Bartok2013}.
 We also introduce an averaging process for the magnetic partial spectra to
 eliminate dependencies on artificial choices of the coordinate origin
 in atomic clusters or crystals and discuss the appropriate methods.

\subsection{Multipole expansion of local magnetic environment}
\label{Sec:MultipoleExp}

 Magnetic structures are usually represented by magnetic moment vectors
 {\red with} a certain order on discretely arranged atoms. 
To obtain the continuous vector function that characterizes the magnetic
 structure, we convert the magnetic moment configurations to the
 magnetization density by introducing the Gaussian function, choosing
 the coordinate's origin at one of the atomic positions as follows:
\begin{align}
{\bm m}({\bm r}) = \sum_{j=1}^{N}e^{-\alpha({\bm r}-{\bm
 R}_{j})^2}{\bm m}_{j}\ ,
\label{Eq:Magvector}
\end{align}
where ${\bm r}$ is spatial position, ${\bm R}_{j}$ is the position
of the $j$-th atom, ${\bm m}_{j}$ is the magnetic moment at the $j$-th
atom, and $\alpha$=$\frac{1}{2\sigma^2}$ with a variance parameter $\sigma$.

The magnetization density is expanded with a set of vector spherical
harmonics ${\bm Y}^{L}_{{\red \ell} m}(\hat{{\bm r}})$ (${\red \ell}\geq
1$, $-{\red \ell}\leq m \leq
{\red \ell}$, $L={\red \ell}-1,{\red \ell},{\red \ell}+1$) and radial functions $\phi_{nl}(r)$ as
\begin{align}
 {\bm m}({\bm r}) = \sum_{nL{\red \ell} m}c_{nL{\red \ell}
 m}\phi_{n{\red \ell}}(r){\bm Y}^{L}_{{\red \ell} m}(\hat{{\bm r}})\ .
\label{Eq:MagExpansion}
\end{align}
The vector spherical harmonics ${\bm Y}^{L}_{{\red \ell} m}$ {\red are}
written with ordinary spherical harmonics $Y_{{\red \ell} m}({\bm r})$ 
and the Clebsh-Gordan coefficients $\braket{{\red \ell}_1 m_1; {\red
\ell}_2 m_2 |{\red \ell}_3 m_3}$ as follows~\cite{Varshalovich1988}:
\begin{align}
 {\bm Y}^{L}_{{\red \ell} m}(\hat{{\bm r}}) =
 \sum_{M=-L}^{L}\sum_{\sigma=-1}^{1}\braket{LM; 1\sigma |{\red \ell}
 m} Y_{LM}(\hat{{\bm r}}){\bm e}_{1\sigma}\ ,
\label{Eq:VecSpherical}
\end{align}
where ${\bm e}_{1\sigma}$ are the spherical unit vectors satisfying the orthogonality relation ${\bm
e}_{1\sigma}^{*}\cdot{\bm e}_{1\sigma'}$=$\delta_{\sigma\sigma'}$ and are expressed
with the Cartesian unit vectors as follows:
\begin{align}
 {\bm e}_{1-1}&=\frac{1}{\sqrt{2}}({\bm e}_{x}-i{\bm e}_{y}) \nonumber
 \\
 {\bm e}_{10}&={\bm e}_{z} 
 \label{Eq:SpheUnitVec} \\
 {\bm e}_{11}&=-\frac{1}{\sqrt{2}}({\bm e}_{x}+i{\bm e}_{y}) \nonumber \ .
\end{align}
Since Clebsh-Gordan coefficients $\braket{{\red \ell}_1 m_1; {\red
\ell}_2 m_2 |{\red \ell}_3 m_3}$
have finite values only if $m_1$+$m_2$=$m_3$, Eq. (\ref{Eq:VecSpherical})
is written as follows:
\begin{align}
 {\bm Y}^{L}_{{\red \ell} m}(\hat{{\bm r}}) =
 \sum_{\sigma=-1}^{1}\braket{L\,m\!-\!\sigma; 1\,\sigma |{\red \ell}\,
m} Y_{Lm-\sigma}(\hat{{\bm r}}){\bm e}_{1\sigma}\ , 
\label{Eq:VecY}
\end{align}
where $\braket{L\,m\!-\!\sigma; 1\,\sigma |{\red \ell}\,m}=0$ for $\mid m-\sigma\mid > L$.
The vector spherical harmonics {\red have} the orthogonality relation
\begin{align}
 \int d\Omega {\bm Y}_{{\red \ell} m}^{L{\red *}}(\widehat{{\bm
 r}})\cdot{\bm Y}_{{\red \ell}' m'}^{L'}(\widehat{{\bm r}})
 =\delta_{LL'}\delta_{{\red \ell\ell'}}\delta_{mm'}\ ,
\label{Eq:OrthogVecY}
\end{align}
{\red from the relations ${\bm e}_{1\sigma}^{*}\cdot{\bm
e}_{1\sigma'}$=$\delta_{\sigma\sigma'}$, $\int d\Omega Y_{\ell
m}^{*}({\bm r})Y_{\ell'm'}({\bm r})=\delta_{\ell\ell'}\delta_{mm'}$, and
the unitary relation
$\sum_{M\sigma}\braket{{\red \ell}\,m\,|LM;1\,\sigma}\braket{LM;1\,\sigma|\ell'\,m'}=\delta_{\ell\ell'}\delta_{mm'}$. Eq. (\ref{Eq:VecSpherical})
is written with the Cartesian bases {\red from Eq.} (\ref{Eq:SpheUnitVec})}, as follows:
\begin{align}
  {\bm Y}^{L}_{{\red \ell} m}(\hat{{\bm r}}) &=
 \sum_{\mu=x,y,z}\sum_{M=-L}^{L}C^{{\red \ell} m}_{LM\mu}Y_{LM}(\hat{{\bm r}}){\bm e}_{\mu}
\end{align}
where
\begin{align}
 C^{{\red \ell} m}_{LMx} &= \frac{1}{\sqrt{2}}(\langle LM;1-1 |
 {\red \ell} m \rangle - \braket{ LM;1 1 | {\red \ell} m }) \nonumber \\
 C^{{\red \ell} m}_{LMy} &= \frac{-i}{\sqrt{2}}(\braket{ LM;1-1 | {\red \ell} m } 
 + \braket{LM;11 | {\red \ell} m } ) \\
 C^{{\red \ell} m}_{LMz} &= \braket{ LM;10 | {\red \ell} m } \nonumber
\end{align}

Some forms of radial functions $\phi_{n\ell}$ are suggested for the formula of SOAP in earlier works~\cite{Bartok2013, Himanen2020}.
 We here adopt the radial functions suggested by Ba{\red rt}{\'o}k {\it et
 al}.~\cite{Bartok2013}, which do not depend on angular momentum ${\red \ell}$,
 $\phi_{n{\red \ell}}=\phi_{n}$, and are orthonormalized in 
 the range (0, $r_{\rm cut}$):
\begin{align}
 \int_{0}^{r_{\rm cut}} \phi_{n}(r)\phi_{n'}(r) r^2 dr =\delta_{nn'}\ .
\label{Eq:Orthoggn}
\end{align}

Considering the relation $e^{-\alpha ({\bm r}+{\bm R}_{j})^{2}}$=$4\pi
e^{-\alpha ({\bm r}^2+{\bm R}_{j}^{2})}\sum_{{\red
\ell}m}\mathcal{B}_{{\red \ell}}(2\alpha
r R_{j})Y_{{\red \ell} m}(\widehat{{\bm r}})Y_{{\red \ell} m}^{*}(\widehat{{\bm R}_{j}})$
and the orthogonality relations of Eqs. (\ref{Eq:OrthogVecY}) and (\ref{Eq:Orthoggn}),
the expansion coefficients $c_{nL{\red \ell} m}$ in Eq. (\ref{Eq:MagExpansion})
are obtained from Eqs. (\ref{Eq:Magvector}) and (\ref{Eq:MagExpansion})
as follows:
\begin{widetext}
 \begin{align}
 c_{nL{\red \ell} m}=4\pi \int_{0}^{r_{\rm cut}}\sum_{j=1}^{N}\{e^{-\alpha(r^2+R_j^2)}\mathcal{B}_{L}(2\alpha rR_j)  
 {\bm Y}_{{\red \ell} m}^{L*}(\hat{{\bm R}_j})\cdot {\bm m}_j \}\phi_{n}(r)r^2dr\ ,
\label{Eq:multipole}
 \end{align}
\end{widetext}
where $\mathcal{B}_{L}$ is the modified Bessel function.

When {\red the $i$-th} atom located at the origin has a finite magnetic moment, the
contribution from the central atom, $c^{(0)}_{nL{\red \ell} m}$, is included in
Eq. (\ref{Eq:multipole}) with $R_{j}=0$.
From the modified Bessel functions $\mathcal{B}_{L}(R_{j}=0)=1$ for $L$=0 and 0 for $L\neq$0 and vector spherical harmonics ${\bm
Y}_{00}^{0}$=${\bm 0}$ and ${\bm Y}_{1m}^{0}$=$\sqrt{\frac{1}{4\pi}}{\bm
e}_{1m}$, we obtain the analytic forms of the contribution from the
central magnetic atom in Eq. (\ref{Eq:multipole}) as:
\begin{align}
 c^{(0)}_{n01m} = \sqrt{4\pi} \int_{0}^{r_{\rm cut}}r^2\{e^{-\alpha r^2} {\bm e}_{1
 m}^{*} \cdot {\bm m}_{i} \}\phi_{n}(r)dr\ .
 \label{Eq:CoefR0SO}
\end{align}

\subsection{Partial spectra of local magnetic environment}
\label{Sec:SpectraSO}

Here, we derive the partial spectra of magnetization density, which are
multiple-dimensional vectors characterizing the corresponding magnetic configuration.
The basic procedure to derive the partial spectra is similar to those
discussed in SOAP~\cite{Bartok2013}, except that we {\red address the
distribution of axial vectors}, but there are some points to be noted.
We define an overlap integral between two magnetization distributions
${\bm m}({\bm r})$ and ${\bm m}'({\bm r})$ as follows:
\begin{align}
  S({\bm m}, {\bm m}') \equiv \int d{\bm r} {\bm m}^{*}({\bm
  r})\cdot {\bm m}'({\bm r}) \ .
\label{Eq:MagOverlap}
\end{align}
The rotational invariants quantifying the similarity of the two
magnetization densities are then obtained as follows:
\begin{align}
 k^{(\xi)}({\bm m}, {\bm m}')=\int d\hat{R} \mid S({\bm m}, \hat{R}{\bm m}')\mid^{\xi}\ ,
\label{Eq:OrigKernel}
\end{align}
where $\hat{R}$ is the rotation operation. {\red We refer to the quantity $k^{(\xi)}({\bm m}, {\bm m}')$
as $\xi$-th order similarity kernel of ${\bm m}$ and ${\bm m}'$.}
 We note that $k^{(\xi)}(\bm m,\bm m')$ in Eq. (\ref{Eq:OrigKernel})
 {\red for odd $\xi$, which may be defined such as $k^{(3)}({\bm m},
{\bm m}')= \int d\hat{R}S^{*}({\bm m},\hat{R}{\bm m}')S({\bm m},\hat{R}{\bm m}')S({\bm
m},\hat{R}{\bm m}')$, is always zero when the 
magnetic structure under consideration has, for instance, a two-fold rotation
$\hat{R}_{2}$ symmetry satisfying $\hat{R}_{2}{\bm m}({\bm r})=-{\bm m}({\bm r})$
 due to $S({\bm m}, \hat{R}_{2}{\bm m}')=-S({\bm m}, {\bm m}')$}.
$\hat{R}{\bm m}$ in Eq.	(\ref{Eq:OrigKernel}) is calculated by using the
transformation relation of the vector spherical harmonics for the rotation operation $R$ as:
\begin{align}
 \hat{R}{\bm Y}^{L}_{{\red \ell} m} = \sum_{m'}{\bm Y}^{L}_{{\red \ell}
 m'}D^{{\red \ell}}_{m'm}(\hat{R})\ ,
\label{Eq:RotationMat}
\end{align}
where {\red $D^{\ell}_{mm'}$ is the matrix elements of the unitary
representation matrix of the rotation operation for vector spherical
harmonics, $D^{\ell}$}, satisfying the relation
\begin{align}
  D^{{\red \ell}\dagger}(\hat{R})D^{{\red \ell}}(\hat{R}) =I\ .
\end{align}
{\red From Eqs. (\ref{Eq:MagExpansion}), (\ref{Eq:OrthogVecY}),
(\ref{Eq:Orthoggn}), and (\ref{Eq:RotationMat}), 
 the overlap integral in Eq. (\ref{Eq:OrigKernel}) is calculated as follows:
\begin{align}
 S({\bm m},\hat{R}{\bm m}') =\sum_{nL\ell}\sum_{mm'}c_{nL\ell
 m}^{*}c_{nL\ell m'}'D_{mm'}^{\ell}(\hat{R})
\label{Eq:CalcMagOverlap}
\end{align}
}

We here introduce the inner product of the multi-dimensional vectors ${\bm
A}$ and ${\bm B}${\red ,} whose complex components are identified by
multiple indices as follows:
\begin{align}
  \braket{{\bm A}, {\bm B}} = \sum_{\mu} A_{\mu}^{*} B_{\mu}\ ,
\label{Eq:GeneralInnerProduct}
\end{align}
where $\mu$ represents all the indices specifying the components.
The second{\red -}order similarity kernel {\red of magnetization densities
${\bm m}$ and ${\bm m}'$} is then obtained as
{\red
\begin{align}
 k^{(2)}({\bm m}, {\bm m}') &= \int d\hat{R}S^{*}({\bm
 m},\hat{R}{\bm m}')S({\bm m},\hat{R}{\bm m}') \nonumber \\
&= \sum_{nn'}\sum_{LL'}\sum_{\ell}(P_{nLn'L'\ell})^{*}P_{nLn'L'\ell}' \nonumber \\
&= \braket{{\bm P}, {\bm P}'}\ ,
\label{Eq:SecondSimilarityKernel}
\end{align}
}
where ${\bm P}$ and ${\bm P}'$ are {\red considered vectors} composed of
the following elements: 
{\red
\begin{align}
  P_{nLn'L'\ell}&=\sqrt{\frac{8\pi^2}{2\ell+1}}\sum_{m} c^{*}_{nL\ell m}c_{n'L'\ell m} \nonumber \\
             &=\sqrt{\frac{8\pi^2}{2\ell+1}}\braket{{\bm c}_{nL\ell},{\bm c}_{n'L'\ell}}\ .
\label{Eq:PowerSO}
\end{align}
}
{\red To derive Eqs. (\ref{Eq:SecondSimilarityKernel}) and
 (\ref{Eq:PowerSO}), we used Eq. (\ref{Eq:CalcMagOverlap}) and the relation
\begin{align}
 \int d\hat{R}D^{\ell_{1}
 *}_{m_{1}m_{1}'}(\hat{R})D^{\ell_2}_{m_{2}m_{2}'}(\hat{R})=\frac{8\pi^{2}}{2\ell_{1}
 +1}\delta_{\ell_{1}\ell_{2}}\delta_{m_{1}m_{2}}\delta_{m_{1}'m_{2}'}\ .
\label{Eq:RelationDInt}
\end{align}
}
 We refer to ${\bm P}$ (${\bm P}'$) as a magnetic power spectrum for
 magnetization density ${\bm m}$ (${\bm m}'$), which is similar to that
 defined for atomic densities~\cite{Bartok2013}.
{\red As discussed after Eq. (\ref{Eq:OrigKernel}), the
 similarity kernel of Eq. (\ref{Eq:OrigKernel}) vanishes for odd $\xi$
 when the magnetic structure has specific symmetry. 
 Therefore, the bispectrum {\red that} can be derived from
 Eq. (\ref{Eq:OrigKernel}) for $\xi$=3, as discussed for SOAP in
 Ref. \onlinecite{Bartok2013}, is not appropriate
 as the descriptor of the magnetic structure.}
The fourth{\red -}order similarity kernel is also derived from
 Eq.~(\ref{Eq:OrigKernel}) as follows:
{\red
\begin{align}
 k^{(4)}({\bm m}, {\bm m}') &=\int d\hat{R}\{S^{*}({\bm
 m},\hat{R}{\bm m}')S({\bm m},\hat{R}{\bm m}')\}^{2}\nonumber \\
 &= \sum_{\gamma_{1}\gamma_{2}\gamma_{3}\gamma_{4}}\sum_{\ell=|\ell_{1}-\ell_{3}|}^{\ell_{1}+\ell_{3}}(T_{\gamma_{1}\gamma_{2}\gamma_{3}\gamma_{4}\ell})^{*}T'_{\gamma_{1}\gamma_{2}\gamma_{3}\gamma_{4}\ell}\nonumber \\
&= \braket{{\bm T}, {\bm T}'}\ ,
\label{Eq:FourthSimilarityKernel}
\end{align}
}
 where the multi-dimensional vector ${\bm T}$ (${\bm T}'$), referred
 {\red to} as
 magnetic {\red trispectrum}, has the vector elements:
{\red
\begin{align}
 T_{\gamma_1\gamma_2\gamma_3\gamma_4\ell}&= \sqrt{\frac{8\pi^2}{2\ell +
 1}}\sum_{m}{\red g_{\gamma_{1}\gamma_{2}\ell
 m}^{*}g_{\gamma_{3}\gamma_{4}\ell m}} \nonumber\\
&=\sqrt{\frac{8\pi^2}{2\ell + 1}}
 \braket{{\bm g}_{\gamma_1\gamma_2\ell},{\bm g}_{\gamma_3\gamma_4\ell}}\ .
\label{Eq:TriSO}
\end{align}
}
The $\gamma_i$ represent{\red s} the set of $n_i$,$L_i$,${\red \ell}_i$,
i.e., $\gamma_i$=$\{n_i,L_i,{\red \ell}_i\}$, and ${\red \ell}$ runs from
$\max\{\mid {\red \ell}_1-{\red \ell}_2 \mid, \mid {\red \ell}_3-{\red \ell}_4\mid\}$ to
$\min\{{\red \ell}_1+{\red \ell}_2, {\red \ell}_3+{\red \ell}_4\}$ and the vector elements of
${\bm g}_{\gamma_1\gamma_2{\red \ell}}$ are
\begin{align}
  {\red g_{\gamma_1\gamma_2\ell m}} = \sum_{m'} c_{\gamma_1
  m'} c_{\gamma_2 m\!-\!m'} \braket{{\red \ell}_1 m';{\red \ell}_2
 m\!-\!m'|{\red \ell} m}\ ,
\label{Eq:gSO}
\end{align}
where $m$ takes integer values from $-{\red \ell}$ to ${\red \ell}$.
{\red To derive Eqs. (\ref{Eq:FourthSimilarityKernel})-(\ref{Eq:gSO}),
we used the relation
\begin{align}
 D^{\ell_{1}}_{m_{1}m_{1}'}(\hat{R}) D^{\ell_{2}}_{m_{2}m_{2}'}(\hat{R})=&\nonumber\\
\sum_{\ell=|\ell_{1}-\ell_{2}|}^{\ell_{1}+\ell_{2}}\sum_{mm'}\braket{\ell_{1}m_{1}\ell_{2}m_{2}|\ell m}&\braket{\ell_{1}m_{1}'\ell_{2}m_{2}'|\ell m'}D^{\ell}_{mm'}(\hat{R})
\label{Eq:RelationDProd}
\end{align}
 in addition to Eqs. (\ref{Eq:CalcMagOverlap}) and
 (\ref{Eq:RelationDInt}). }

{\red The name "trispectrum" for the quantity of Eq. (\ref{Eq:TriSO}) is
taken after a terminology in signal theory, as similar to "power spectrum" and
"bispectrum"~\cite{Collis1998}.}
As discussed in Sec. \ref{Sec:SymmetryTest}, a higher{\red -}order partial
spectrum is necessary to distinguish the magnetic structures given
by the same atomic configuration, for instance, magnetic
structures only with different magnetic anisotropies.

\subsection{Modified partial spectrum irrespective of magnetic anisotropy}
\label{Sec:SpectraNSO}
{\red In the derivation of the partial spectra in the previous
subsections, it was assumed that the magnetic moments rotate in
accordance with the spatial rotation of the crystal. The coupling
between crystal axes and magnetic moment is realized through the spin-orbit
interaction, which is a relativistic effect. Meanwhile,
f}irst-principles calculations for {\red magnetic systems} are often
implemented without taking spin-orbit interaction to make the
calculations faster or to understand the magnetic states with the
simplified picture of the spin space. 
{\red In the absence of spin-orbit interaction, the rotation of magnetic
moments and crystal axes can be performed independently.
 As a result, magnetic structures arising from rotations in
spin have the same total energy.}

 It is thus useful to introduce the modified partial spectrum representation,
which does not distinguish the magnetic structures only with different
magnetic anisotropy.
 To obtain such a partial spectrum, we expand the magnetization density
 with the product function of the normal spherical harmonics and the
 unit vector as the bases of the classical spin space {\red as}
\begin{align}
 {\bm m}({\bm r}) = \sum_{n{\red \ell} m\sigma}\overline{c}_{n{\red \ell} m\sigma}\phi_{n}(r)
 Y_{{\red \ell} m}(\hat{{\bm r}}){\bm e}_{1\sigma}
\ .
\label{Eq:magnetizationSpin}
\end{align}
From Eqs. (\ref{Eq:Magvector}) and (\ref{Eq:magnetizationSpin}), the
expansion coefficients $\overline{c}_{n{\red \ell} m\sigma}$ are calculated as follows:
\begin{widetext}
  \begin{align}
  \overline{c}_{n{\red \ell} m\sigma}=4\pi \int_{0}^{r_{\rm cut}}\sum_{j}^{N}\{e^{-\alpha(r^2+R_j^2)}
  \mathcal{B}_{{\red \ell}}(2\alpha rR_j) Y_{{\red \ell} m}^{*}(\hat{{\bm R}_j}) {\bm e}^{*}_{1\sigma} \cdot 
  {\bm m}_j \}\phi_{n}(r)r^2dr\ .
 \label{Eq:multipoleNSO}
  \end{align}
 \end{widetext}
The contribution from the {\red $i$-th magnetic atom located at the origin} with $R_j=0$,
$\overline{c}^{(0)}_{n{\red \ell} m\sigma}$, in
Eq.(\ref{Eq:multipoleNSO}) is finite only for ${\red \ell}$=$m$=$0$ and has the
analytic form as follows:
\begin{align}
 \overline{c}^{(0)}_{n00\sigma} = \sqrt{4\pi} \int_{0}^{r_{\rm cut}}r^2\{e^{-\alpha r^2} {\bm e}_{1
 \sigma}^{*} \cdot {\bm m}_{i} \}\phi_{n}(r)dr\ .
 \label{Eq:CoefR0NSO}
\end{align}

The magnetic structures only with different magnetic anisotropy are
transformed with each other through a rotation of the spin moments only in
the spin space.
 Therefore, the partial spectra to characterize magnetic moment
 configurations irrespective of magnetic anisotropy are given by the
 following similarity kernel:
\begin{align}
 \overline{k}^{(\xi)}({\bm m}, {\bm m}')=\int d{\red \hat{R}} \int
 d{\red \hat{R}_s} \mid S({\bm m}, {\red \hat{R}\hat{R}_s}{\bm
 m}')\mid^{\xi}\ ,
\label{Eq:OrigKernelNSO}
\end{align}
where the rotation operator for magnetization density now works
separately for the spatial coordinate ($R$) and spin coordinate ($R_{s}$) as follows:
\begin{align}
  {\red \hat{R}\hat{R}_{s}}{\bm m}({\bm r}) = \sum_{n{\red \ell}
 m\sigma} \overline{c}_{n{\red \ell} m\sigma}\phi_{n}(r)
 \sum_{m'\sigma'}Y_{{\red \ell} m'}{\bm e}_{1\sigma'}D^{{\red \ell}}_{m'm}(\hat{R})D^{s}_{\sigma'\sigma}(\hat{R}_s) .
\end{align}
Eq. (\ref{Eq:OrigKernelNSO}) leads to the power spectrum for $\xi$=2 as:
\begin{align}
 \overline{P}_{nn'{\red \ell}}=\sqrt{\frac{8\pi^2}{2{\red \ell} +
 1}}\sqrt{\frac{8\pi^2}{3}} \braket{\overline{{\bm c}}_{n{\red \ell}},
 \overline{{\bm c}}_{n'{\red \ell}}} \ ,
\end{align}
and the trispectrum for $\xi$=4:
\begin{align}
 \overline{T}_{n_1{\red \ell}_1n_2{\red \ell}_2n_3{\red \ell}_3n_4{\red
 \ell}_4}^{{\red \ell} j} =
 \sqrt{\frac{8\pi^2}{2{\red \ell} +
 1}}\sqrt{\frac{8\pi^2}{2j+1}}\braket{\overline{{\bm g}}_{n_1{\red \ell}_1
 n_2{\red \ell}_2}^{{\red \ell} j}, \overline{{\bm g}}_{n_3{\red \ell}_3
 n_4{\red \ell}_4}^{{\red \ell} j}}\ ,
\end{align}
where
\begin{widetext}
 
\begin{align}
 {\red \overline{g}^{\ell j}_{n_1\ell_1 n_2\ell_2 m\sigma}} =
 \sum_{m'=-{\red \ell}}^{{\red
 \ell}}\sum_{\sigma'=-j}^{j}\overline{c}_{n{\red \ell}
 m'\sigma'}\overline{c}_{n'{\red \ell}'m\!-\!m'\sigma\!-\!\sigma'}
 \braket{{\red \ell}_1
 m';{\red \ell}_2 m\!-\!m'|{\red \ell} m}\braket{1 \sigma'; 1 \sigma\!-\!\sigma'|j \sigma}\ ,
\label{Eq:gvecNSO}
\end{align}
\end{widetext}
where {\red $\ell$} runs from $|{\red \ell}_1\!-\!{\red \ell}_2|$ to
${\red \ell}_1\!+\!{\red \ell}_2$ and $j$ from 0 to 2.

\subsection{Elimination of origin choice dependency}
\label{Sec:Averaging}
 As a descriptor, the magnetic partial spectrum should not depend on the
 artificial choices of the origin of the coordinate in the atomic system. To eliminate the origin choice dependency of
 the magnetic partial spectra in Sec. \ref{Sec:SpectraSO} and \ref{Sec:SpectraNSO}, we redefine the
 partial spectra by taking the average of the expansion coefficients over
 origin choices at all atoms in the atomic cluster {\red or the crystal's unit cell}.
The average of atomic positions can be applied directly for the
expansion coefficients ${\bm c}^{i}_{nL{\red \ell}}${\red , where $i$ indicates the
 atomic site chosen as the coordinate's origin}, similar to the partial
power spectrum for atomic positions~\cite{Bartok2013}, such as
\begin{align}
 P_{nLn'L'{\red \ell}}^{({\rm in})} &=
 \sqrt{\frac{8\pi^2}{2{\red
 \ell}+1}}\frac{1}{N^2}\braket{\biggl(\sum_{i=1}^{N}{\bm c}^{i}_{nL{\red
 \ell}}\biggr),
 \biggl(\sum_{j=1}^{N}{\bm c}^{j}_{n'L'{\red \ell}}\biggr)}\ ,
 \label{Eq:PowerInner}\\
  T_{\gamma_1\gamma_2\gamma_3\gamma_4{\red \ell}}^{({\rm in})} &=
 \sqrt{\frac{8\pi^2}{2{\red \ell} + 1}}\frac{1}{N^2}\braket{(\sum_{i=1}^{N} {\bm
 g}_{\gamma_1\gamma_2{\red \ell}}^{i}),
(\sum_{j=1}^{N}{\bm g}^{j}_{\gamma_3\gamma_4{\red \ell}})}\ {\red ,}
 \label{Eq:TriInner}
\end{align} 
{\red where $N$ is the number of atoms in the atomic cluster or
the crystal's unit cell.}
 {\red One o}ther way of origin-choice average over the atomic positions for the
product of the expansion coefficients is given as follows:
\begin{align}
 P_{nLn'L'{\red \ell}}^{({\rm out})} &=
 \sqrt{\frac{8\pi^2}{2{\red
 \ell}+1}}\frac{1}{N}\sum_{i=1}^{N}\braket{{\bm c}^{i}_{nL{\red
 \ell}},{\bm c}^{i}_{n'L'{\red \ell}}}\ ,
\label{Eq:PowerOuter}\\
  T_{\gamma_1\gamma_2\gamma_3\gamma_4{\red \ell}}^{({\rm out})} &=
 \sqrt{\frac{8\pi^2}{2{\red \ell} + 1}} \frac{1}{N} 
  \sum_{i=1}^{N} \braket{{\bm g}_{\gamma_1\gamma_2{\red \ell}}^{i},{\bm
 g}^{i}_{\gamma_3\gamma_4{\red \ell}}}\ .
\label{Eq:TriOuter}
\end{align}
We refer to the average implemented in Eq. (\ref{Eq:PowerInner}) and
(\ref{Eq:TriInner}) as {\red the} inner average and those in
Eq. (\ref{Eq:PowerOuter}) and Eq. (\ref{Eq:TriOuter}) as {\red the} outer average.

The inner average for the power spectrum can be zero for specific
magnetic structures, such as the {\red typical} antiferromagnetic configuration whose
magnetic moments are alternating on each sub-lattice.
Hereafter, we take the outer average for the power spectrum
and the inner average for {\red the} trispectrum. {\red This redefines the
second- and fourth-order similarity kernels with the power spectrum ${\bm P}$ and trispectrum ${\bm T}$ as follows:
\begin{align}
  k^{(2)}({\bm m}, {\bm m}') &= \sum_{nn'}\sum_{LL'}\sum_{\ell}(P_{nLn'L'\ell}^{({\rm out})})^{*}P_{nLn'L'\ell}^{\prime({\rm out})} \nonumber \\
&= \braket{{\bm P}, {\bm P}'}\ ,\\
 k^{(4)}({\bm m}, {\bm m}')  &=
 \sum_{\gamma_{1}\gamma_{2}\gamma_{3}\gamma_{4}}\sum_{\ell=|\ell_{1}-\ell_{3}|}^{\ell_{1}+\ell_{3}}(T_{\gamma_{1}\gamma_{2}\gamma_{3}\gamma_{4}\ell}^{({\rm
 in})})^{*}T_{\gamma_{1}\gamma_{2}\gamma_{3}\gamma_{4}\ell}^{\prime({\rm
 in})}\nonumber \\
&= \braket{{\bm T}, {\bm T}'}
\end{align}
}

As discussed in {\red Sec.~\ref{Sec:NumericalTest}}, the power spectrum ${\bm P}$
is insufficient to fully distinguish magnetic structures with distinct
magnetic symmetries in the same atomic configuration, and a higher{\red -}order
spectrum is necessary to distinguish such magnetic structures through
the overlap of magnetization densities at different atomic sites.


\section{Numerical results}
\label{Sec:NumericalTest}

In this section, we demonstrate the parameter dependence and
discrimination performance of the magnetic partial spectra derived in
Sec. \ref{Sec:Formulation} for the magnetic structures in high{\red -}symmetry crystals.
  To measure the similarity of magnetization density, we use the normalized
 similarity kernels defined as follows:
{\red
\begin{align}
K^{(2)}({\bm m},{\bm m}')=\frac{\braket{{\bm P}, {\bm P}'}}{
\sqrt{\braket{{\bm P},{\bm P}}}
\sqrt{\braket{{\bm P}',{\bm P}'}}}\ ,\label{eq:k2}\\
K^{(4)}{\red ({\bm m},{\bm m}')}=\frac{\braket{{\bm T}, {\bm T}'}}{\sqrt{\braket{{\bm T}, {\bm T}}}\sqrt{\braket{{\bm T}',{\bm T}'}}}\ ,\label{eq:k4}
\end{align}
where ${\bm P}$ (${\bm P}'$) and ${\bm T}$ (${\bm T}'$) are {\red the} power
spectrum and trispectrum for the magnetization density ${\bm m}$ (${\bm m}'$). 
We also define the normalized similarity kernels irrespective of
magnetic anisotropy by using $\overline{\bm P}$, $\overline{\bm P}'$,
$\overline{\bm T}$, $\overline{\bm T}'$, derived in
Sec.~\ref{Sec:SpectraNSO}, instead of ${\bm P}$, ${\bm P}'$, ${\bm T}$,
${\bm T}'$ in Eqs. (\ref{eq:k2}) and (\ref{eq:k4}) as
$\overline{K}^{(2)}({\bm m},{\bm m}')$ and $\overline{K}^{(4)}({\bm
m},{\bm m}')$, respectively.
}

These similarity kernels take the positive value in the range $[0,1]$ 
since the similarity kernel is derived from Eq. (\ref{Eq:OrigKernel}) or
(\ref{Eq:OrigKernelNSO}) with even $\xi$.
We have implemented the calculations of SOAP with magnetic alignment by
modifying the {\red P}ython library, DScribe~\cite{Himanen2020}.

\subsection{Parameter dependence of the similarity kernels}
\label{Sec:ParameterTest}
\begin{figure}[t]
 \includegraphics[width=0.8\linewidth]{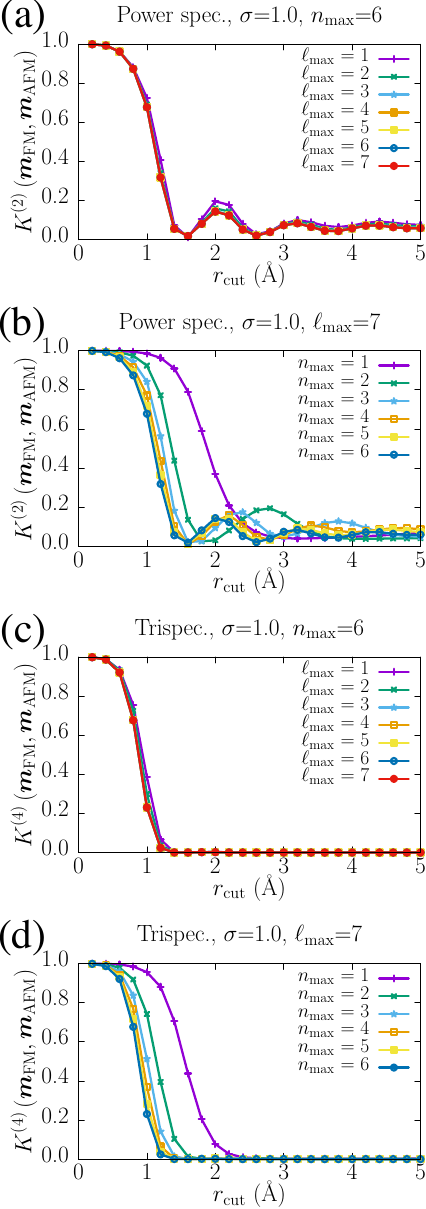}
 \caption{
The $r_{\rm cut}$ dependence of the similarity kernels for the
 ferromagnetic (FM) and antiferromagnetic (AFM)
 systems. Panels (a)-(d) show the $K^{(2)}$ and $K^{(4)}$ values as a
 function of 
 ${\red \ell}_{\rm max}$ with fixed $n_{\rm max}=6$ 
 or 
 $n_{\rm max}$ with fixed ${\red \ell}_{\rm max}=7$. 
 The magnetic moments are ordered along the $z$-axis
 on each atomic site in the one-dimensional periodic lattice with 1 \AA\
interval along the $x$-axis.
}
\label{Fig:ParamDep1Dim}
\end{figure}
In this section, we discuss how the similarity kernel $K^{(2)}$ and
$K^{(4)}$, defined by Eq.~\eqref{eq:k2} with the power spectrum $\bm P$ and
\eqref{eq:k4} with the trispectrum $\bm T$, respectively, behave in
{\red the} application for magnetic configurations on simple crystals. 
Note that $K^{(2)}$ and $K^{(4)}$ contain four parameters: the width of the
magnetization distribution $\sigma$ ({\it i.e.}, $\alpha$), the number of radial basis
functions $n_{\rm max}$, the maximum angular momentum of the vector
spherical harmonics ${\red \ell}_{\rm max}$, and $r_{\rm cut}$ as the cutoff for the
radial integration in Eq
.~\eqref{Eq:multipole}. We will discuss these parameter dependences of
$K^{(2)}$ and $K^{(4)}$ and show that $n_{\rm max}$ and ${\red \ell}_{\rm max}$ need to be
sufficiently large to achieve convergence, while appropriate values for
$\sigma$ and $r_{\rm cut}$ must be chosen according to {\red an} individual purpose.

Let us first discuss the simplest case, a one-dimensional chain with spins placed
at intervals of 1 \AA\ along the $x$-axis, and consider the similarity
kernel between the FM and AFM ordered systems. 
In Fig.~\ref{Fig:ParamDep1Dim}, we show the similarity
kernels $K^{(2)}$ and $K^{(4)}$. Here, we fix the two pairs of
parameters, namely, $\sigma$ and $n_{\rm max}$ ((a) and (c)), and $\sigma$
and ${\red \ell}_{\rm max}$ ((b) and (d)), and change other parameters. 
In all cases,
$K^{(2)}$ and $K^{(4)}$ converge as $n_{\rm max}$ or ${\red \ell}_{\rm max}$ become
larger. The convergence is faster for ${\red \ell}_{\rm max}$ than for
$n_{\rm max}$ since
we are {\red working} with a one-dimensional chain where angular dependence is
less significant. On the other hand, $K^{(2)}$ and $K^{(4)}$ change 
from 1 to 0 as $r_{\rm cut}$ increases. This can be understood from
Eq.~\eqref{Eq:multipole}, which suggests that {\red a} small $r_{\rm cut}$ including
only one spin in the region cannot distinguish the FM and AFM states. 
Here, $K^{(2)}$ and $K^{(4)}$ show almost the same $r_{\rm cut}$ dependence,
with the only difference being that $K^{(4)}$ approaches zero faster than $K^{(2)}$.

\begin{figure}[t]
 \includegraphics[width=0.8\linewidth]{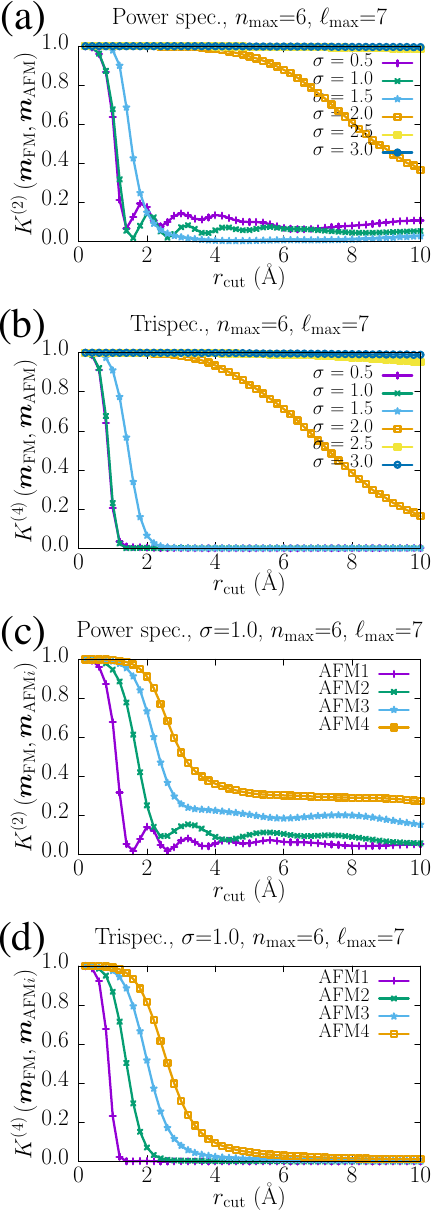}
 \caption{
 Similarity kernels with different values of $\sigma$, using $n_{\rm
 max}$=6 and ${\red \ell}_{\rm max}$=7 for FM and AFM configu{\red
 r}ations on the one{\red -}dimensional crystal, (a) $K^{(2)}$ and (b) $K^{(4)}$. 
Panels (c) and (d) compare the similarity kernels for FM and different
 AFM$i$ magnetic patterns, where the magnetic moments have opposite signs
 per $i$-atomic sites. 
$\sigma = 1.0$\AA, $n_{\rm max}$=6, and ${\red \ell}_{\rm max}$=7 are used.
}
\label{Fig:SigDep1Dim}
\end{figure}

Figures \ref{Fig:SigDep1Dim}(a) and (b) show the $\sigma$ and $r_{\rm cut}$ dependence of $K^{(2)}$
and $K^{(4)}$, respectively, with converged values of
${\red \ell}_{\rm max}$ and $n_{\rm max}$. We can see that increasing $\sigma$ shifts the
crossover point between the FM and AFM states to larger values of
$r_{\rm cut}$. In the large $\sigma$ limit, $K^{(2)}\sim K^{(4)} \sim 1$
regardless of the value of $r_{\rm cut}$. This is reasonable with
Eqs.~\eqref{Eq:multipole}, as well as
Eqs.~\eqref{eq:k2} and ~\eqref{eq:k4}. Namely, $e^{-\alpha(\bm r+\bm
R_j)^2}$ factor becomes constant in the large $\sigma$ limit, resulting
in a constant factor change depending on the magnetic structure. As this constant
change is lost in the normalization in Eqs.~\eqref{eq:k2} and
~\eqref{eq:k4}, the resulting $K$ shows no distinction between FM and
AFM configurations. These findings imply that the appropriate choice of
$\sigma$ and $r_{\rm cut}$ is needed for the magnetic partial spectra to get
meaningful information about the magnetic structure. 
To get more insight, we show {\red the normalized similarity kernels,}
$K^{(2)}$ and $K^{(4)}${\red , between the FM order and several
AFM orders} in Figs.~\ref{Fig:SigDep1Dim} (c) and (d). Here,
AFM1, AFM2, AFM3, and AFM4 correspond to $\uparrow\downarrow$,
$\uparrow\uparrow\downarrow\downarrow$,
$\uparrow\uparrow\uparrow\downarrow\downarrow\downarrow$, and
$\uparrow\uparrow\uparrow\uparrow\downarrow\downarrow\downarrow\downarrow$
periodic magnetic structures, respectively, in the one{\red -}dimensional
crystal.
 Obviously, AFM4 is closer to FM than AFM3, while AFM1 is the farthest
 from FM, indicating that accurate discrimination among these states is
 crucial when comparing similarity with FM. 
When we use $\sigma$=1.0\AA, as shown in Fig.~\ref{Fig:SigDep1Dim}(c) and
(d), the four states can be distinguished by using $r_{\rm cut}\sim 3$ with
$K^{(2)}$. In contrast, with trispectrum, $K^{(4)}$,
the differences between the four states are relatively small compared to
$K^{(2)}$ for large $r_{\rm cut}$, but they can still be distinguishable.
 The results indicate that the power spectrum is a better descriptor than
the {\red trispectrum} for this specific purpose, as it allows for more apparent
discrimination of these magnetic configurations.



\begin{figure}[t]
\includegraphics[width=0.8\linewidth]{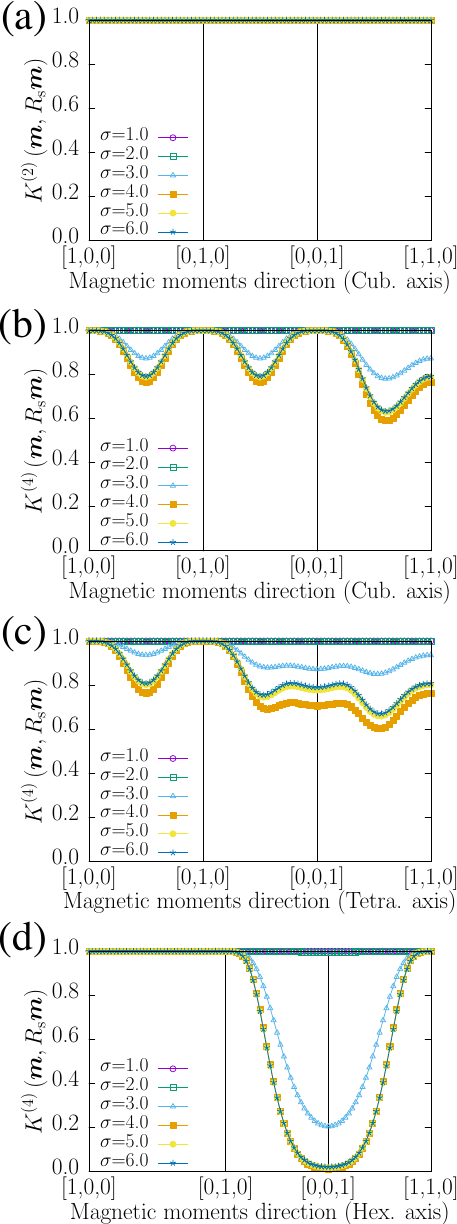}
\caption{
The spin rotation dependence of the normalized similarity kernels
 between the magnetization density of the ferromagnetic structure along
 the (100)-axis, denoted by ${\bm m}$, and those of the rotated magnetic
 moments, given by $R_{s}(\theta,\phi){\bm m}$. 
 The results are presented for (a)$K^{(2)}$ and (b) $K^{(4)}$ on a
 simple BCC lattice with lattice constant $a$=1.0\AA, (c) $K^{(4)}$ on a
 body center tetragonal lattice with $a$=1.0\AA\ and $c$=1.5\AA, and (d)
 $K^{(4)}$ on a hexagonal lattice with $a$=1.0\AA\ and $c$=1.5\AA, using
 $n_{\rm max}$=4, ${\red \ell}_{\rm max}$=6, and $r_{\rm cut}$=5.0\AA.
}
\label{Fig:anisomag}
\end{figure}

Based on these results, we propose the following strategies for
determining appropriate values of $r_{\rm cut}$ and $\sigma$:
\begin{itemize}
\item First, set $r_{\rm cut}$ depending on the purpose. For
      example, when classifying the magnetic structures, $r_{\rm cut}$
      should be chosen based on their typical spatial scale to be
      distinguished. 
\item Then, $\sigma$ should be chosen so that the most distinct typical
      magnetic structures realized within the chosen $r_{\rm cut}$, such as
      the conventional FM and AFM states, approach $K\sim0$.
\end{itemize}
Up to now, $K^{(2)}$ and $K^{(4)}$ do not show a significant difference. 
However, for the purpose {\red of distinguishing} the magnetic structures with
different magnetic symmetries, we will show that $K^{(4)}$ would be a
better descriptor than $K^{(2)}$.

Figure \ref{Fig:anisomag} displays the angle dependence of the similarity
kernels between simple ferromagnetic orders along the $x$-axis and
different directions of magnetic moments in 
polar coordinates
$(\theta,\phi)$ on a simple body center cubic, tetragonal, and hexagonal
lattices with their origin $(\theta=0,\phi=0)$ set to the $x$-axis. In 
all crystal systems, the crystal $a$-axis is set {\red to 1 \AA,} and the
$c$-axis of tetragonal and hexagonal lattices is set to 1.5\AA.
 The plots show that, while the second{\red -}order similarity kernels $K^{(2)}$ do not
show any difference for the magnetic anisotropy, the similarity kernel
$K^{(4)}$ captures the difference from magnetic anisotropy. 
Note that the difference is captured through the overlap of the
magnetization density around neighboring atoms since some extent of
$\sigma$ is required to capture the difference of magnetic anisotropy,
as shown in Fig.~\ref{Fig:anisomag}.
 {\red As shown in the following subsection, $K^{(4)}$ has better discrimination
performance than $K^{(2)}$ for magnetic anisotropy, though $K^{(2)}$ is not entirely useless for differentiating
magnetic structures that differ only in magnetic anisotropy.}

Note that the modified partial spectra insensitive to magnetic
anisotropy are designed not to reflect the anisotropy difference even
for the higher{\red -}order partial spectra. As a result, $\overline{K}^{(2)}$
and $\overline{K}^{(4)}$ both show 1.0 for any ($\theta$, $\phi$) as
well as the plots of $K^{(2)}$ in Fig. \ref{Fig:anisomag} (a).
The difference of the magnetic anisotropy for the rotation along
the $z$-axis for the hexagonal lattice is not reflected even for the
fourth{\red -}order partial spectrum $K^{(4)}$, as shown in Fig.~\ref{Fig:anisomag} (d),
implying that the magnetic partial spectrum with the order higher than four is
necessary to capture such a difference.
{\red Differentiating} magnetic anisotropy is essential for
classifying magnetic symmetries {\red of the} magnetic alignments {\red on} the
same atomic configurations, as discussed in Sec.~\ref{Sec:SymmetryTest}.

\subsection{Discrimination performance for magnetic structures with distinct magnetic symmetries}
\label{Sec:SymmetryTest}
Magnetic symmetries of the magnetic systems determine whether or not various
physical properties of magnetic materials occur, such as anomalous Hall
and Nernst effect, electromagnetic effect{\red ,} and magnetic Kerr effect.
The ability to distinguish different magnetic symmetries of a descriptor
is thus crucial to analyze the physical properties of magnetic materials
by using machine learning.

\begin{figure}[h]
\includegraphics[width=1.0\linewidth]{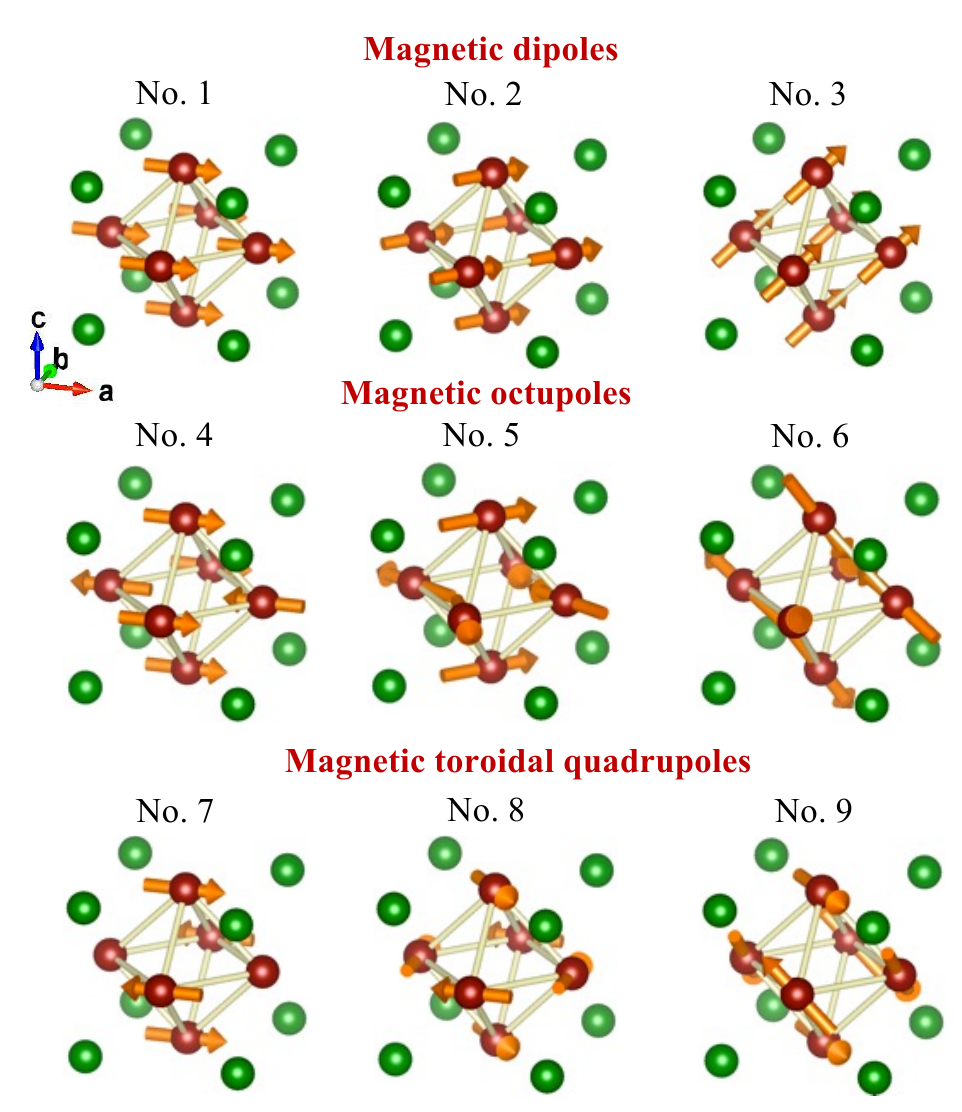}
\caption{Symmetrized magnetic structures generated on
 Mn$_3$Ir crystal by cluster multipole method with
magnetic moment normalized to unity \cite{MTS2019}.}
\label{Fig:MultipolesMn3Ir}
\end{figure}

\begin{figure*}[h]
  \includegraphics[width=1.0\linewidth]{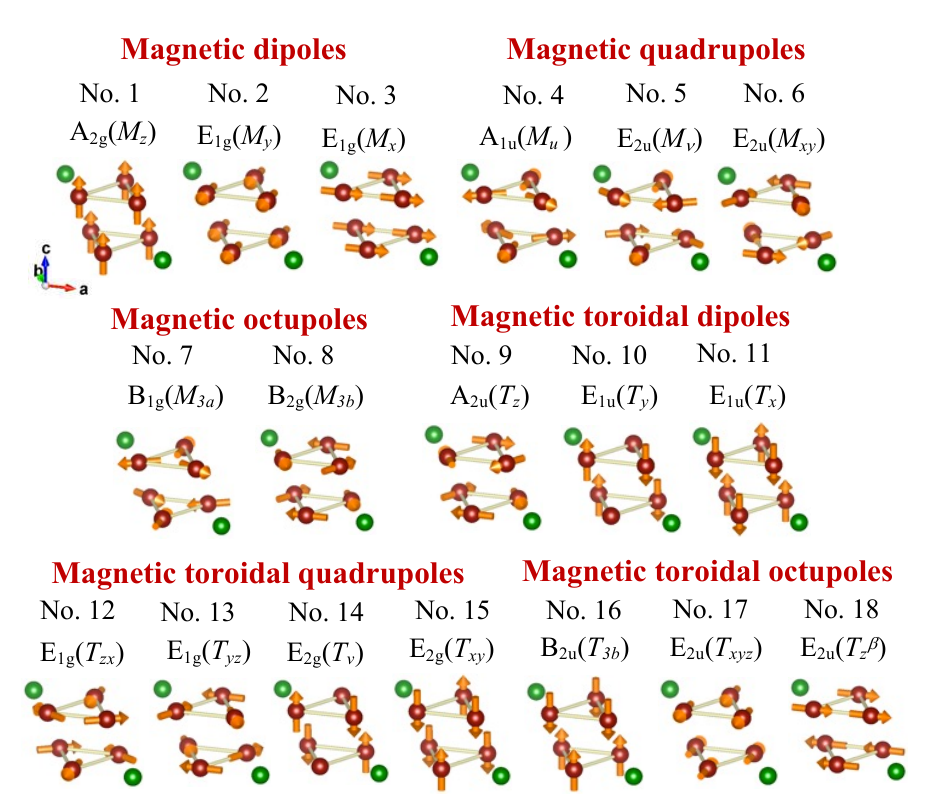}
  \caption{Symmetrized magnetic structures generated on
 Mn$_3$Sn crystal by cluster multipole method with
 each finite magnetic moment normalized to unity~\cite{MTS2019}.}
\label{Fig:MultipolesMn3Sn}
\end{figure*}

Although the expansion coefficients $c_{nLlm}$ of the vector spherical harmonics
for magnetization density in Eq. (\ref{Eq:MagExpansion}) contain all
information {\red on} magnetic environments, some information may be lost through
the procedure of constructing the partial spectra from the expansion coefficients.
To examine the ability to capture magnetic symmetry in the current scheme, 
we investigate the similarity kernels for the magnetic structures
classified according to the symmetries of magnetic structures in high
symmetry crystals, cubic Mn$_3$Ir and hexagonal Mn$_3$Sn.
The crystal structures of Mn$_3$Ir and Mn$_3$Sn {\red belong} to the space group
$Pm\overline{3}m$ ($O_{h}^{1}$, No.221), and $P6_{3}/mmc$
($D_{6h}^{4}$, No.194), respectively. The lattice constants are $a$=3.77
\AA\ for Mn$_3$Ir and $a$=5.665\AA\ and $c$=4.531\AA\ for
Mn$_3$Sn.
The magnetic structures of Mn$_3$Ir and Mn$_3$Sn, which are classified
according to the irreducible representations of their respective crystal
point groups ($O_h$ for cubic Mn$_3$Ir and $D_{6h}$ for hexagonal
Mn$_3$Sn), were generated using the cluster multipole method described
in Ref.~\onlinecite{MTS2019}. 
{\red With the method, the magnetic structure bases classified by multipoles
symmetrized according to the irreducible representations of the
crystallographic point group are systematically generated by first generating magnetic structures on
virtual atomic clusters belonging to the point group that
conforms to the multipole moments symmetrized according to the point
group, and then mapping them onto atoms in the crystal in a way that
preserves convertibility with respect to the point group operations~\cite{MTS2019}.}
For Mn$_3$Ir, the magnetic structures with different magnetic symmetries
within the same multipoles are {\red produced} by taking the linear
combination of the generated magnetic bases as explained in
Ref. \onlinecite{Huyen2019}.
In each magnetic structure, the size of the magnetic moment was
normalized to one at each magnetic site. The symmetrized magnetic
structures for cubic Mn$_3$Ir and hexagonal Mn$_3$Sn are shown in
Fig.\ref{Fig:MultipolesMn3Ir} and Fig.\ref{Fig:MultipolesMn3Sn},
respectively. Additional details regarding the crystal and magnetic
structures can be found in the Supplementary Information.

 {\red Several potential approaches can be considered for applying this
 method to crystals comprising various atom species. Here, we will take the
 simplest approach and treat only the magnetic atoms of interest,
 the Mn atoms, and ignore the other atoms, implying that
 the presence of nonmagnetic atoms affects the formation of the magnetic
 structure even though it is not treated explicitly.
 This method makes it possible to compare the crystals composed of different crystal
 structures and atomic species as demonstrated by comparing the magnetic
 structures of Mn$_3$Ir and Mn$_3$Sn later.}

\begin{figure*}[b]
\includegraphics[width=1.0\linewidth]{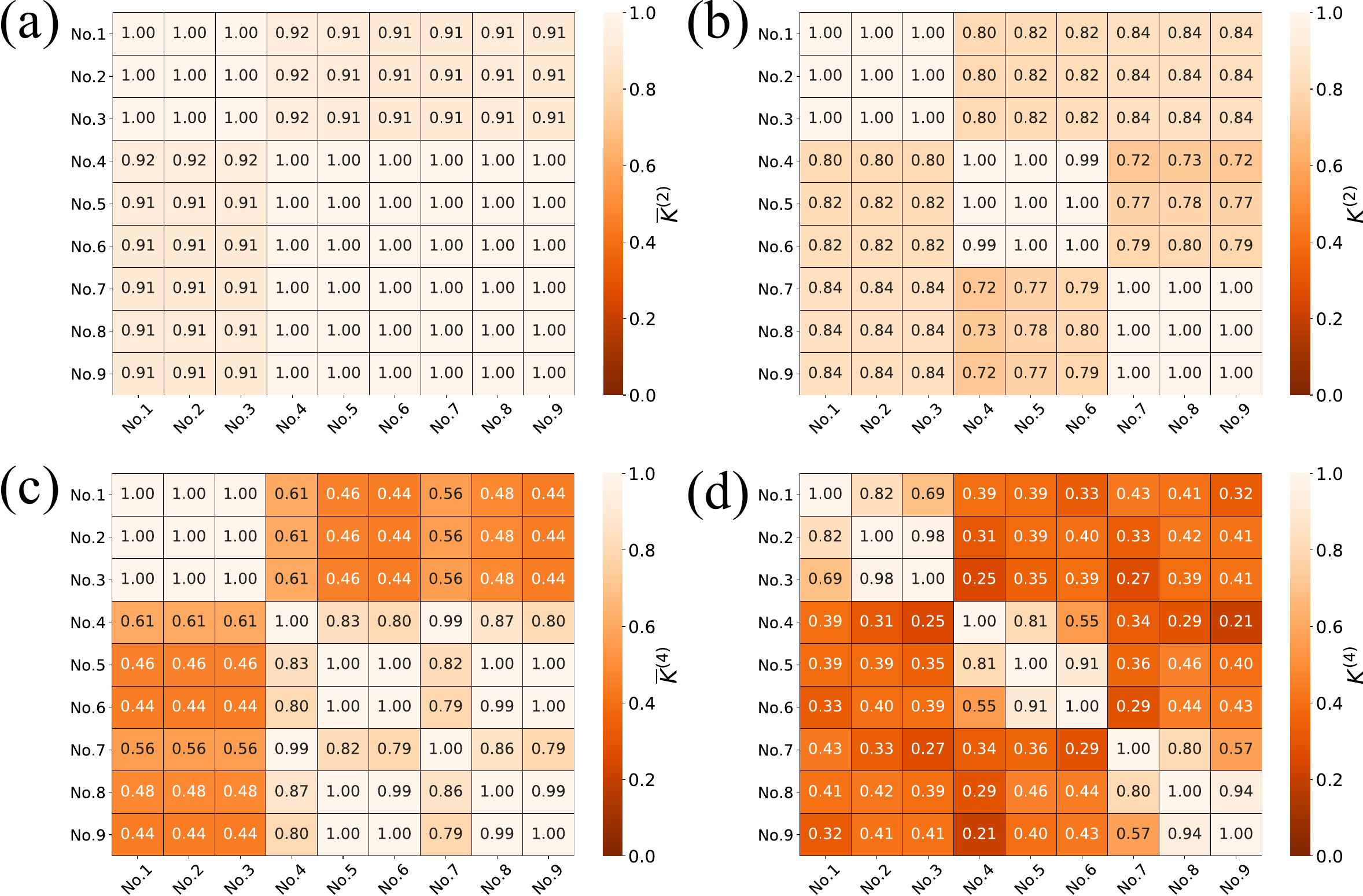}
\caption{Pair plots of {\red normalized similarity kernels} of {\red the
 power spectrum} and {\red trispectrum} with
 $\sigma$=4.0\AA, $n_{\rm max}$=4, ${\red \ell}_{\rm max}$=6, and $r_{\rm
 cut}$=5\AA\ for the magnetic structures corresponding to
 Fig. \ref{Fig:MultipolesMn3Ir} for (a) $\overline{K}^{(2)}$, (b) $K^{(2)}$,
 (c) $\overline{K}^{(4)}$, and (d) $K^{(4)}$.}
\label{Fig:CorrelationMn3Ir}
\end{figure*}

\begin{figure*}[b]
    \includegraphics[width=1.0\linewidth]{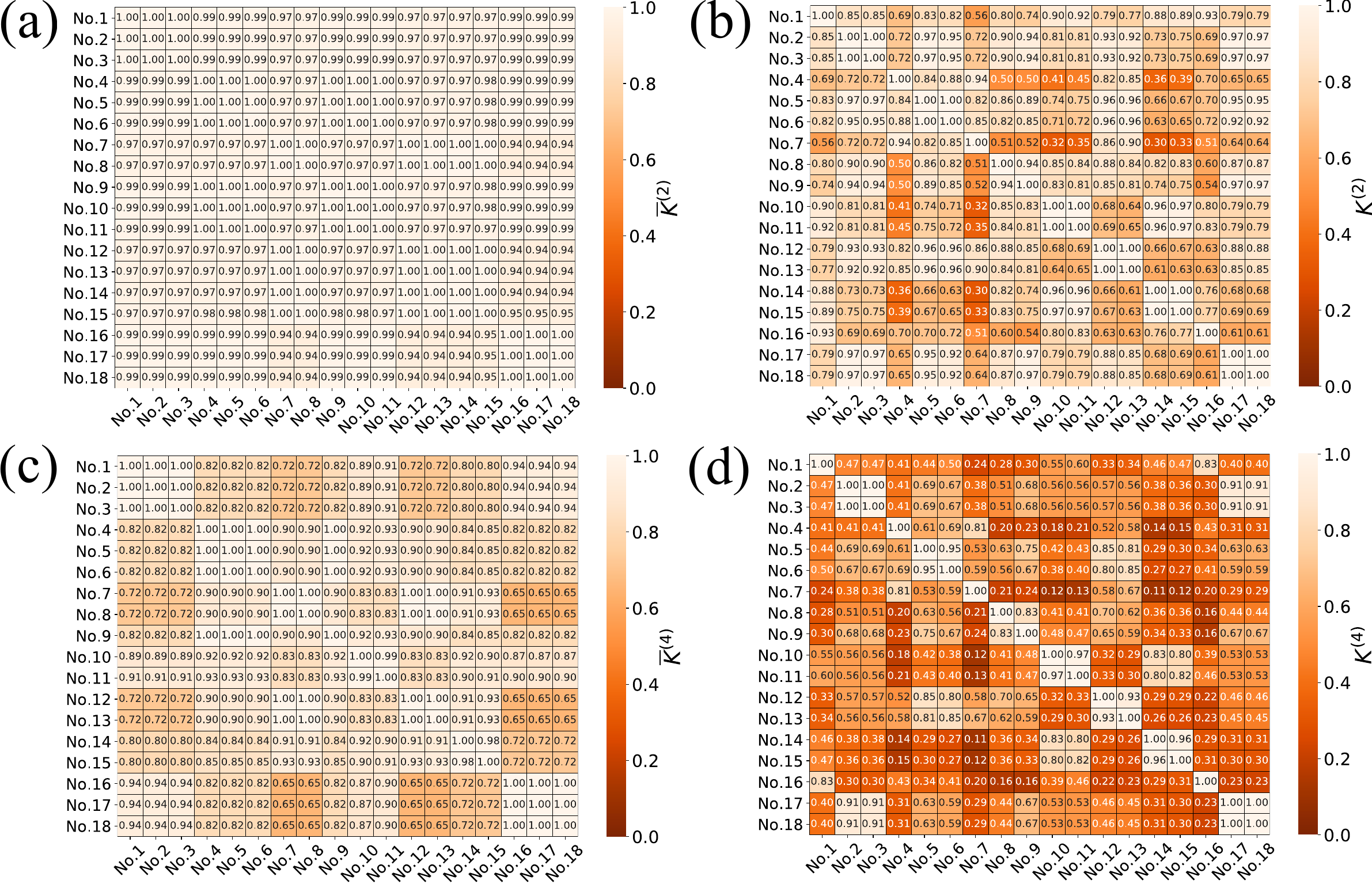}
    \caption{Pair plots of {\red normalized similarity kernels} of {\red
 the power spectrum} and {\red trispectrum} with
     $\sigma$=4.0\AA, $n_{\rm max}$=4, ${\red \ell}_{\rm max}$=6, and $r_{\rm
 cut}$=5\AA\ for magnetic structures corresponding to
 Fig.~\ref{Fig:MultipolesMn3Sn} for (a) $\overline{K}^{(2)}$, (b)
 $K^{(2)}$, (c) $\overline{K}^{(4)}$, and (d) $K^{(4)}$.}
  \label{Fig:CorrelationMn3Sn}
\end{figure*}

Figures \ref{Fig:CorrelationMn3Ir} and \ref{Fig:CorrelationMn3Sn} show
the correlation tables {\red of} the {\red normalized} similarity ke{\red r}nels
$\overline{K}^{(2)}$, $K^{(2)}$, $\overline{K}^{(4)}$, {\red and}
$K^{(4)}$ {\red calculated} for the symmetrized magnetic structures of
Fig.~\ref{Fig:MultipolesMn3Ir} and \ref{Fig:MultipolesMn3Sn}, {\red respectively.
 The} similarity kernels $K^{(\xi)}$ ($\xi=2, 4$) show better resolution
to distinguish the different magnetic structures than those designed to
neglect the magnetic anisotropy, $\overline{K}^{(\xi)}$, for the same $\xi$,
and the fourth{\red -}order similarity kernels show better resolution than the
second{\red -}order ones. Thus, $\overline{K}^{(2)}$ has the least ability to
distinguish the magnetic structure{\red ,} while $K^{(4)}$ demonstrate{\red s the} maximum
capability to distinguish the different magnetic structures{\red ,} as discussed
in detail below.

The magnetic dipole structures of No. 1-3 in
Figs.~\ref{Fig:MultipolesMn3Ir} and~\ref{Fig:MultipolesMn3Sn} are
ordinary ferromagnetic structures along different axes and {\red differ} only in magnetic anisotropy.
In Fig.~\ref{Fig:MultipolesMn3Ir}, since the magnetic structures No. 5
and No. 6 are obtained by 90 degrees {\red of} spin
rotation on each atom of No. 8 and No. 9, respectively,
those two magnetic structures differ only in magnetic anisotropy. 
Similarly, in Fig.~\ref{Fig:MultipolesMn3Sn}, the magnetic
structures of No. 5 and No. 7 differ {\red from No. 6 and
No. 8}, respectively, {\red in} magnetic anisotropy.
As a result, the $\overline{K}^{(\xi)}$ evaluate that they are
equivalent for those magnetic structures irrespective of the order
$\xi$, as shown in Figs.~\ref{Fig:CorrelationMn3Ir} and
\ref{Fig:CorrelationMn3Sn} for $\xi$=2, 4.
In addition, $\overline{K}^{(2)}$ and $\overline{K}^{(4)}$ barely reflect the magnetic structural
differences that go beyond the magnetic anisotropy, such as No. 5 and 6 in
Fig.~\ref{Fig:MultipolesMn3Ir} and No. 4 and 5 in
Fig.~\ref{Fig:MultipolesMn3Sn}, by showing the value
close to 1.0.
 $K^{(2)}$ shows moderate discrimination performance for magnetic
 structures with different magnetic symmetries and succeeds {\red in finding}
 differences in magnetic anisotropy, but it still fails to discriminate
 several magnetic structures that involve differences in magnetic anisotropy.
$K^{(4)}$ shows significantly better discrimination performance
for magnetic structures {\red distinct from} both magnetic symmetry and magnetic anisotropy.
 However, even with the $K^{(4)}$, it is hard to distinguish the
 difference in the hexagonal in-plane magnetic anisotropy between No.2
 and 3 and between No.17 and 18{\red ,} as discussed in Sec. \ref{Sec:ParameterTest},
suggesting a partial spectrum of the order higher than four is required
to distinguish the difference of the hexagonal in-plane magnetic anisotropy.


\begin{figure*}[b]
  \includegraphics[width=1.0\linewidth]{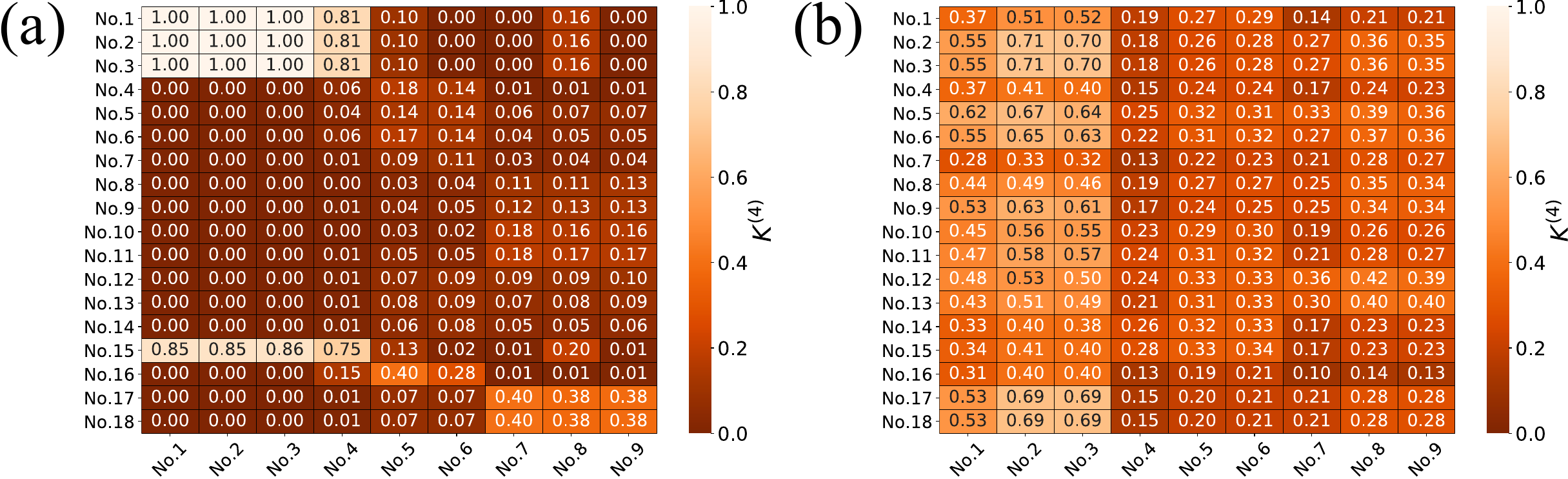}
  \caption{Pair plots of {\red normalized similarity kernel} $K^{(4)}$ between
   the symmetrized magnetic structures of Mn$_3$Ir and of Mn$_3$Sn,
 shown in Fig.~\ref{Fig:MultipolesMn3Ir} and
 Fig.~\ref{Fig:MultipolesMn3Sn}. The calculations use $n_{\rm max}$=4,
 ${\red \ell}_{\rm max}$=6,
 and $r_{\rm cut}$=5\AA\ with (a) $\sigma$=1.0\AA\ and (b) $\sigma$=4.0\AA.}
\label{Fig:CorrelationMn3SnMn3Ir}
\end{figure*}

{\red As mentioned above, t}he similarity of the local magnetic environment can be evaluated for different 
magnetic compounds with different crystal structures by focusing on specific
magnetic ions that both compounds contain.
Figure~\ref{Fig:CorrelationMn3SnMn3Ir} provides the correlation table between 
the magnetic structures of Mn$_3$Ir in Fig.~\ref{Fig:MultipolesMn3Ir}
and those of Mn$_3$Sn in Fig.~\ref{Fig:MultipolesMn3Sn}, using
$\sigma$=1.0\AA\ and $\sigma$=4.0\AA\ with $n_{\rm max}$=4,
${\red \ell}_{\rm max}$=6, and $r_{\rm cut}$=5.0\AA.
For $\sigma$=1.0\AA, the width of the magnetization density around each Mn
atom is small, and there is little overlap between the
magnetization densities coming from different magnetic atoms,
resulting in the dominant contribution of Eq.~\eqref{Eq:CoefR0SO} for the
{\red trispectrum}, Eqs.~\eqref{Eq:TriInner} and \eqref{Eq:gSO}.
In this case, the difference {\red in} the magnetic structures is reflected
only through the averaging procedure of the origin choice in
Eq.~\eqref{Eq:TriInner}, and the ferromagnetic structures can not be
distinguished even for the different crystals. In
Fig.~\ref{Fig:CorrelationMn3SnMn3Ir}, No. 4 of Mn$_3$Ir and No. 15
of Mn$_3$Sn are evaluated as closer to ferromagnetism than the others. 
This reflects that these magnetic structures have finite
magnetizations in their ferrimagnetic structure.
The magnetization densities broadened with a value of
$\sigma$=4.0\AA\ lead to significant overlap between the
magnetization densities from different magnetic atoms.
This causes the difference in the partial spectra for the distinct
magnetic environments around the Mn atoms in Mn$_3$Ir and Mn$_3$Sn,
{\red increasing} discrimination performance for both
compounds even in the case of ferromagnetic structures, as depicted in
Fig.~\ref{Fig:CorrelationMn3SnMn3Ir} (b).

\section{Summary}
\label{Sec:Summary}
We have developed a theory of descriptors for magnetic structures {\red
invariant} to arbitrary choices of the coordinate axis, crystal unit cells, and
origin choices. Higher-order partial spectra are essential to discriminate
different magnetic structures that share the same atomic
positions, especially with difference{\red s} in magnetic anisotropy. 
We have also derived {\red the} fourth-order partial spectr{\red um}, referred to as the
{\red trispectrum}, and compared its properties {\red to the} second-order
partial spectrum, i.e., the power spectrum. 
The {\red trispectrum} effectively distinguishes magnetic structures only with 
magnetic anisotropy difference, though the power spectrum cannot discriminate them.
 Therefore, the {\red trispectrum} is necessary
{\red to accurately classify magnetic symmetries relevant to the} physical
properties of magnetic materials.
Additionally, we have derived alternative partial spectra {\red irrelevant} to
magnetic anisotropy and confirmed their effectiveness. 
These modified partial spectra are particularly useful for
classifying magnetic structures obtained from first-principles
calculations without spin-orbit coupling. 
Our theory of descriptors for magnetic structures thus provides a
powerful tool for accurately classifying magnetic structures and paves
the way for new applications in materials science and engineering by
machine learning.

\section*{Acknowledgments}
This research is supported by JSPS KAKENHI Grants Numbers JP19H01842,
 JP19K03752, JP20H05262, JP20K05299, JP20K21067, JP21H01789, JP21H04437, JP21H01031, and
 JP22H00290, JP23K03288, JP23H00091, and by JST PRESTO Grant Number JPMJPR17N8 and JPMJPR20L7.
 We also acknowledge the use of supercomputing system, MASAMUNE-IMR, at
 CCMS, IMR, Tohoku University in Japan.

\bibliographystyle{apsrev}
\bibliography{SOAPmag}

\clearpage
\widetext

\section*{Supplementary information}
The crystal of Mn$_3$Ir belongs to the space group $Pm\overline{3}m$ ($O_{h}^{1}$,
No.221) and has three Mn atoms on the $3c$ Wycoff site [Mn1:(0,$\frac{1}{2}$,$\frac{1}{2}$), Mn2: ($\frac{1}{2}$,0,$\frac{1}{2}$), Mn3:
($\frac{1}{2}$,$\frac{1}{2}$,0)] and one Ir atom on the $1a$ site at (0,0,0)
in the unit cell. 
Mn$_3$Sn belongs to the space group $P6_{3}/mmc$ ($D_{6h}^{4}$, No.194)
and has six Mn atoms on $6h$ site [Mn1:($x$,2$x$,$\frac{1}{4}$), Mn2:($\overline{x}$,$x$,$\frac{3}{4}$), Mn3:($\overline{x}$,2$\overline{x}$,$\frac{3}{4}$), 
Mn4:($x$,$\overline{x}$,$\frac{1}{4}$),
Mn5:(2$\overline{x}$,$\overline{x}$,$\frac{1}{4}$),Mn6:(2$x$,$x$,$\frac{3}{4}$) with $x$=0.8388] 
and Sn atoms on $2d$ site [Sn1:($\frac{1}{3}$, $\frac{2}{3}$, $\frac{3}{4}$), Sn2:($\frac{2}{3}$,$\frac{1}{3}$,$\frac{1}{4}$)].
Magnetic moment at each Mn site of Mn$_3$Ir and Mn$_3$Sn are provided in Tabs. \ref{Tab:MagnStructMn3Ir} and \ref{Tab:MagnStructMn3Sn}.

\renewcommand{\thetable}{S\arabic{table}}

\begin{center}
 \begin{table*}[b]
  \caption{List of symmetrized magnetic structures in Mn$_3$Ir.}
\begin{tabular}{cccc}
No. 1&Mn1:( 1.000000, 0.000000, 0.000000)&Mn2:( 1.000000, 0.000000, 0.000000)&Mn3:( 1.000000, 0.000000, 0.000000)\\
No. 2&Mn1:( 0.707107, 0.707107, 0.000000)&Mn2:( 0.707107, 0.707107, 0.000000)&Mn3:( 0.707107, 0.707107, 0.000000)\\
No. 3&Mn1:( 0.577350, 0.577350, 0.577350)&Mn2:( 0.577350, 0.577350, 0.577350)&Mn3:( 0.577350, 0.577350, 0.577350)\\
No. 4&Mn1:(-1.000000, 0.000000, 0.000000)&Mn2:( 1.000000, 0.000000, 0.000000)&Mn3:( 1.000000, 0.000000, 0.000000)\\
No. 5&Mn1:(-0.894427, 0.447214, 0.000000)&Mn2:( 0.447214,-0.894427, 0.000000)&Mn3:( 0.707107, 0.707107, 0.000000)\\
No. 6&Mn1:(-0.816497, 0.408248, 0.408248)&Mn2:( 0.408248,-0.816497, 0.408248)&Mn3:( 0.408248, 0.408248,-0.816497)\\
No. 7&Mn1:( 0.000000, 0.000000, 0.000000)&Mn2:(-1.000000, 0.000000, 0.000000)&Mn3:( 1.000000, 0.000000, 0.000000)\\
Mo. 8&Mn1:( 0.000000, 1.000000, 0.000000)&Mn2:(-1.000000, 0.000000, 0.000000)&Mn3:( 0.707107,-0.707107, 0.000000)\\
No. 9&Mn1:( 0.000000, 0.707107,-0.707107)&Mn2:(-0.707107, 0.000000, 0.707107)&Mn3:( 0.707107,-0.707107, 0.000000)
\end{tabular}
\label{Tab:MagnStructMn3Ir}
  \caption{List of symmetrized magnetic structures in Mn$_3$Sn.}
\begin{tabular}{ccccccc}
No. 1&Mn1:( 0.000000  0.000000  1.000000)&Mn2:( 0.000000  0.000000  1.000000)&Mn3:( 0.000000  0.000000  1.000000)\\
     &Mn4:( 0.000000  0.000000  1.000000)&Mn5:( 0.000000  0.000000  1.000000)&Mn6:( 0.000000  0.000000  1.000000)\\
No. 2&Mn1:( 0.000000,-1.000000, 0.000000)&Mn2:( 0.000000,-1.000000, 0.000000)&Mn3:( 0.000000,-1.000000, 0.000000)\\
     &Mn4:( 0.000000,-1.000000, 0.000000)&Mn5:( 0.000000,-1.000000, 0.000000)&Mn6:( 0.000000,-1.000000, 0.000000)\\
No. 3&Mn1:( 1.000000, 0.000000, 0.000000)&Mn2:( 1.000000, 0.000000, 0.000000)&Mn3:( 1.000000, 0.000000, 0.000000)\\
     &Mn4:( 1.000000, 0.000000, 0.000000)&Mn5:( 1.000000, 0.000000, 0.000000)&Mn6:( 1.000000, 0.000000, 0.000000)\\
No. 4&Mn1:( 0.000000,-1.000000, 0.000000)&Mn2:( 0.866025,-0.500000, 0.000000)&Mn3:( 0.000000, 1.000000, 0.000000)\\
     &Mn4:(-0.866025, 0.500000, 0.000000)&Mn5:( 0.866025, 0.500000, 0.000000)&Mn6:(-0.866025,-0.500000, 0.000000)\\
No. 5&Mn1:( 0.000000,-1.000000, 0.000000)&Mn2:(-0.866025,-0.500000, 0.000000)&Mn3:( 0.000000, 1.000000, 0.000000)\\
     &Mn4:( 0.866025, 0.500000, 0.000000)&Mn5:(-0.866025, 0.500000, 0.000000)&Mn6:( 0.866025,-0.500000, 0.000000)\\
No. 6&Mn1:( 1.000000, 0.000000, 0.000000)&Mn2:( 0.500000,-0.866025, 0.000000)&Mn3:(-1.000000, 0.000000, 0.000000)\\
     &Mn4:(-0.500000, 0.866025, 0.000000)&Mn5:(-0.500000,-0.866025, 0.000000)&Mn6:( 0.500000, 0.866025, 0.000000)\\
No. 7&Mn1:( 0.000000, 1.000000, 0.000000)&Mn2:( 0.866025,-0.500000, 0.000000)&Mn3:( 0.000000, 1.000000, 0.000000)\\
     &Mn4:( 0.866025,-0.500000, 0.000000)&Mn5:(-0.866025,-0.500000, 0.000000)&Mn6:(-0.866025,-0.500000, 0.000000)\\
Mo. 8&Mn1:(-1.000000, 0.000000, 0.000000)&Mn2:( 0.500000, 0.866025, 0.000000)&Mn3:(-1.000000, 0.000000, 0.000000)\\
     &Mn4:( 0.500000, 0.866025, 0.000000)&Mn5:( 0.500000,-0.866025, 0.000000)&Mn6:( 0.500000,-0.866025, 0.000000)\\
No. 9&Mn1:(-1.000000, 0.000000, 0.000000)&Mn2:(-0.500000,-0.866025, 0.000000)&Mn3:( 1.000000, 0.000000, 0.000000)\\
     &Mn4:( 0.500000, 0.866025, 0.000000)&Mn5:( 0.500000,-0.866025, 0.000000)&Mn6:(-0.500000, 0.866025, 0.000000)\\
No.10&Mn1:( 0.000000, 0.000000, 0.000000)&Mn2:( 0.000000, 0.000000,-1.000000)&Mn3:( 0.000000, 0.000000, 0.000000)\\
     &Mn4:( 0.000000, 0.000000, 1.000000)&Mn5:( 0.000000, 0.000000,-1.000000)&Mn6:( 0.000000, 0.000000, 1.000000)\\
No.11&Mn1:( 0.000000, 0.000000,-1.000000)&Mn2:( 0.000000, 0.000000,-1.000000)&Mn3:( 0.000000, 0.000000, 1.000000)\\
     &Mn4:( 0.000000, 0.000000, 1.000000)&Mn5:( 0.000000, 0.000000, 1.000000)&Mn6:( 0.000000, 0.000000,-1.000000)\\
No.12&Mn1:( 0.000000,-1.000000, 0.000000)&Mn2:( 0.866025, 0.500000, 0.000000)&Mn3:( 0.000000,-1.000000, 0.000000)\\
     &Mn4:( 0.866025, 0.500000, 0.000000)&Mn5:(-0.866025, 0.500000, 0.000000)&Mn6:(-0.866025, 0.500000, 0.000000)\\
No.13&Mn1:(-1.000000, 0.000000, 0.000000)&Mn2:( 0.500000,-0.866025, 0.000000)&Mn3:(-1.000000, 0.000000, 0.000000)\\
     &Mn4:( 0.500000,-0.866025, 0.000000)&Mn5:( 0.500000, 0.866025, 0.000000)&Mn6:( 0.500000, 0.866025, 0.000000)\\
No.14&Mn1:( 0.000000, 0.000000, 0.000000)&Mn2:( 0.000000, 0.000000,-1.000000)&Mn3:( 0.000000, 0.000000, 0.000000)\\
     &Mn4:( 0.000000, 0.000000,-1.000000)&Mn5:( 0.000000, 0.000000, 1.000000)&Mn6:( 0.000000, 0.000000, 1.000000)\\
No.15&Mn1:( 0.000000, 0.000000, 1.000000)&Mn2:( 0.000000, 0.000000,-1.000000)&Mn3:( 0.000000, 0.000000, 1.000000)\\
     &Mn4:( 0.000000, 0.000000,-1.000000)&Mn5:( 0.000000, 0.000000,-1.000000)&Mn6:( 0.000000, 0.000000,-1.000000)\\
No.16&Mn1:( 0.000000, 0.000000, 1.000000)&Mn2:( 0.000000, 0.000000,-1.000000)&Mn3:( 0.000000, 0.000000,-1.000000)\\
     &Mn4:( 0.000000, 0.000000, 1.000000)&Mn5:( 0.000000, 0.000000, 1.000000)&Mn6:( 0.000000, 0.000000,-1.000000)\\
No.17&Mn1:( 0.000000, 1.000000, 0.000000)&Mn2:( 0.000000,-1.000000, 0.000000)&Mn3:( 0.000000,-1.000000, 0.000000)\\
     &Mn4:( 0.000000, 1.000000, 0.000000)&Mn5:( 0.000000, 1.000000, 0.000000)&Mn6:( 0.000000,-1.000000, 0.000000)\\
No.18&Mn1:(-1.000000, 0.000000, 0.000000)&Mn2:( 1.000000, 0.000000, 0.000000)&Mn3:( 1.000000, 0.000000, 0.000000)\\
     &Mn4:(-1.000000, 0.000000, 0.000000)&Mn5:(-1.000000, 0.000000, 0.000000)&Mn6:( 1.000000, 0.000000, 0.000000)
\end{tabular}
\label{Tab:MagnStructMn3Sn}
 \end{table*}
\end{center}

\end{document}